\documentclass[superscriptaddress,twocolumn,showpacs,amsmath,amssymb,prl,floatfix]{revtex4}

\usepackage{graphicx}
\usepackage{dcolumn}
\usepackage{bm}
\usepackage{bbm}
\usepackage{psfrag}
\usepackage{multirow,amssymb,amsbsy,amsmath}
\usepackage{stmaryrd}

\newcommand{\dd}{\text{d}}
\newcommand{\dod}[2]{\frac{\dd #1}{\dd #2}}

\newcommand{\ddim}{\udelta\kern0.1em}

\newcommand{\beikonst}[2]{\left( #1 \right)_{\kern-0.2em #2}}
\newcommand{\tr}[2][]{\text{Tr}_{#1}\left\{#2\right\}}

\newcommand*{\bra}[1]{\mathopen{\langle}#1\mathclose{|}}
\newcommand*{\ket}[1]{\mathopen{|}#1\mathclose{\rangle}}

\newcommand{\anticom}[2]{\left\{#1,#2\right\}}
\newcommand{\rs}[1]{{\rm \scriptscriptstyle #1}}

\newcommand{\dop}{\hat{\rho}}

\newcommand{\p}{\hat{P}}
\newcommand{\h}{\hat{H}}
\newcommand{\s}{\hat{\sigma}}

\newcommand{\aop}[1]{\hat{A}_{#1}}
\newcommand{\bop}[1]{\hat{A}_{#1}}

\begin{document}

\preprint{APS/123-QED}

%
%

\title{Quantum critical behavior in strongly interacting Rydberg gases}

\author{Hendrik\ Weimer}%
\affiliation{Institute of Theoretical Physics III, Universit\"at Stuttgart, %
              70550 Stuttgart, Germany}%
\email{hweimer@itp3.uni-stuttgart.de}%
\author{Robert\ L\"ow}%
\affiliation{5.~Physikalisches Institut, Universit\"at Stuttgart, %
              70550 Stuttgart, Germany}
\author{Tilman\ Pfau}%
\affiliation{5.~Physikalisches Institut, Universit\"at Stuttgart, %
              70550 Stuttgart, Germany}
\author{Hans\ Peter\ B\"uchler}%
\affiliation{Institute of Theoretical Physics III, Universit\"at Stuttgart, %
              70550 Stuttgart, Germany}%

\date{\today}%

\begin{abstract}

We study the appearance of correlated many-body phenomena in an
ensemble of atoms driven resonantly into a strongly interacting
Rydberg state. The ground state of the Hamiltonian describing the
driven system exhibits a second order quantum phase transition. We
derive the critical theory for the quantum phase transition and show
that it describes the properties of the driven Rydberg system in the
saturated regime. We find that the suppression of Rydberg excitations
known as blockade phenomena exhibits an algebraic scaling law with a
universal exponent.

\end{abstract}


\maketitle

The concept of universality is a powerful tool for the understanding
and characterization of complex phenomena in different fields of
physics. The most pronounced example represents the universal scaling
in systems close to a second order phase transition and the
characterization of the transition in terms of critical theories and
universality classes \cite{Fisher1998}.  Its main assertion is that in
the presence of a diverging length scale, physical observables become
independent on the precise microscopic realization of the systems, and
allows for a qualitative understanding without the knowledge of the
exact microscopic details. In this Letter, we analyze the appearance
of correlated manybody effects in a dense ensemble of atoms driven
resonantly in a strongly interacting Rydberg state, and show that the
behavior of the system can be understood in terms of a critical
theory.

Strongly interacting Rydberg atoms are an area of intense experimental
investigation: the resonant diffusion of Rydberg excitations via the
dipole-dipole interactions has been reported \cite{Anderson1998}, and
the reduction of Rydberg excitation due to blockade effects has been
observed
\cite{Tong2004,Singer2004,Vogt2006,Heidemann2007,Raitzsch2008,Heidemann2008}.
Furthermore, coherent optical excitation has been observed for
individual Rydberg atoms \cite{Johnson2008}, essentially
non-interacting ensembles
\cite{Cubel2005,Mohapatra2007,Reetz-Lamour2008}, as well as in the
strongly interacting regime \cite{Raitzsch2008}.  Of special interest
for the theoretical analysis are cold atomic samples, where the
thermal motion of the atoms is essentially frozen on the
characteristic time scale of the Rydberg excitation (``frozen Rydberg
gas'') \cite{Anderson1998}.  Then, the time evolution of such an
atomic ensemble driven by a resonant Rydberg excitation has been
extensively studied in the past using a numerical integration of the
time-dependent Schr\"odinger equation for small sample sizes
\cite{Robicheaux2005,Stanojevic2008} and within a master equation
approach \cite{Ates2007,Stanojevic2008a}.

\begin{figure}[ht]
  \includegraphics[width= 0.95\columnwidth]{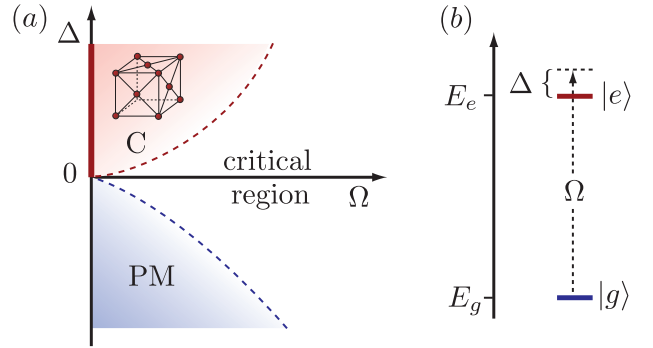}
  \caption{(color online). (a) Phase diagram in the
    $\Delta$-$\Omega$--plane: for a coupling $\Omega = 0$ a second
    order phase transition appears from a crystalline phase (C) to a
    paramagnetic phase (PM), while at a detuning $\Delta=0$ the system
    is dominated by the critical region. (b) Setup of the system with
    $|g\rangle$ the atomic ground state and $|e\rangle$ the excited
    Rydberg state coupled by driving lasers with Rabi frequency
    $\Omega$ and detuning $\Delta$.}  \label{fig:phase}
\end{figure}

In contrast to the previous analyses studying the time evolution, the
approach for the study of strongly interacting Rydberg gases presented
in this Letter is based on the observation that the driven system
relaxes into an equilibrium state. This equilibrium state is dominated
by the thermodynamic phases of the Hamiltonian describing the driven
system, which exhibits a continuous quantum phase transition in the
detuning $\Delta$; see Fig.~\ref{fig:phase}(a). Then, the resonant
regime with $\Delta = 0$ is determined by the critical properties of
the phase transition.  We derive the universal exponents within
mean-field theory, and show that the time evolution of the system and
its relaxation into the equilibrium state is well described by a
master equation derived from the microscopic Hamiltonian.

We start with the Hamiltonian describing an ultracold gas of atoms
driven into an excited Rydberg state with a repulsive van der Waals
interaction. The relevant internal structure for each atom is given by
the atomic ground state $|g\rangle_{i}$ and the excited Rydberg state
$|e\rangle_{i}$, reducing the internal structure to a two-level
system.  The two internal states are coherently coupled by external
lasers with the Rabi frequency $\Omega$ and a detuning $\Delta$; see
Fig.~\ref{fig:phase}(b).  The characteristic time scale for a Rydberg
excitation is short compared to the thermal motion of each atom, and
the positions ${\bf r}_{i}$ of the atoms are frozen
\cite{Anderson1998}: the positions ${\bf r}_{i}$ are randomly
distributed according to the distribution function of a thermal
gas. After performing the rotating wave approximation, the Hamiltonian
in the rotating frame 
can be written as a spin Hamiltonian (cf.  \cite{Robicheaux2005})
\begin{equation}
  \label{eq:H} \h = -\frac{\Delta}{2}\sum\limits_i \s_z^{(i)} +
\frac{\hbar\Omega}{2} \sum\limits_i \s_x^{(i)} +
C_{6} \sum\limits_{j<i}
\frac{\hat{P}_{ee}^{(i)}\hat{P}_{ee}^{(j)}}{|{\bf r}_i-{\bf r}_j|^6},
\end{equation}
where $\hat{P}_{ee}^{(i)}=\ket{e}_i\bra{e}_i= (1+\s_z^{(i)})/2$ 
is the projector onto the excited Rydberg state of the
$i^{\mathrm{th}}$ atom. The last term accounts for the strong van der
Waals repulsion between the Rydberg states with $C_{6}\propto
\bar{n}^{11}$, where $\bar{n}$ is the principle quantum number of the
Rydberg excitation \cite{Gallagher1994,Singer2005}. It is well known
that dense samples of Rydberg atoms undergo collective ionization
processes that might harm the coherent evolution
\cite{Robicheaux2005a}. However for repulsive Rydberg S states these
ionization processes are slowed down significantly
\cite{Amthor2007}. Recent experiments in very dense samples allowed to
follow the coherent evolution involving Rydberg S states on timescales
comparable to the timescales required to reach equilibrium
\cite{Heidemann2007,Raitzsch2008}. In these experiments the saturation
is reached well before radiative or motional decoherence effects set
in. Therefore the validity of the Hamiltonian (\ref{eq:H}) is well
justified for an experimentally realistic situation.

Within the standard experimental setup, the system is prepared with
all the atoms in the ground state and the time evolution is studied by
turning on the excitation lasers, i.e., $|\psi\rangle = \prod_{i}
|g\rangle_{i}$.  While in absence of interactions, the time evolution
of the Hamiltonian (\ref{eq:H}) results in coherent Rabi oscillations
for each atom with the reduced density matrix for each atom being a
pure state, the interactions leads to correlations between different
atoms, and eventually to a decoherence of this single atom pure state.
Consequently, the density matrix describing a subsystem of the atomic
cloud will equilibrate into a steady state under time evolution as the
surrounding atomic states act as a reservoir interacting with the
subsystem.  This behavior is well confirmed by the exact numerical
integration of the system with the Hamiltonian (\ref{eq:H}), see
below, and is also observed in experiments via the saturation of the
Rydberg excitation \cite{Heidemann2007}.  This observation opens up an
alternative approach to study the long-time behavior of the system by
focusing on the equilibrium states of the Hamiltonian (\ref{eq:H}),
and especially on its zero temperature phase diagram.

For vanishing Rabi frequency $\Omega = 0$, the exact ground state of
the Hamiltonian (\ref{eq:H}) can be analytically determined in any
dimension. It is dominated by a second order quantum phase transition
for the critical detuning $\Delta_{c}=0$; see
Fig.~\ref{fig:phase}. For negative detuning $\Delta \leq0$ the ground
state is paramagnetic with all atoms in the atomic state
$|g\rangle_{i}$; i.e., the experimentally relevant initial state
$|\psi\rangle$ is the ground state of the system.  In turn for
positive detuning $\Delta >0$, the ground state prefers to excite
atoms into the Rydberg state and the configuration minimizing the
repulsive van der Waals interaction is obtained for a crystalline
arrangement of the atoms. It is important to notice, that within the
critical region the system is independent on the microscopic
realization, and therefore, its properties are isotropic and
homogeneous.

The behavior of the system for finite $\Omega$ in the resonant regime
$\Delta=0$ is dominated by the critical behavior of the second order
quantum phase transition. At the critical point with $\Delta=0$, the
system is characterized by a single dimensionless parameter $\alpha =
\hbar \Omega/C_{6}n^2$ describing the ratio between the coupling
energy $\hbar \Omega$ and the interaction energy $C_6 n^2$ (here, $n$
denotes the atomic density). Then, the ground state properties such as
the fraction of excited Rydberg atoms exhibit an algebraic behavior
\begin{equation}
  f_{R} \equiv 
  \langle P_{ee}^{(i)} \rangle= c \alpha^{\nu},
\end{equation}
with a universal scaling exponent $\nu$ in the
critical region with $\alpha \ll 1$. In the following, we determine the critical
exponent within mean-field theory.

The important quantity in the mean-field theory is the averaged Rydberg fraction
$f_{R}=\langle P_{ee}^{(i)}\rangle$. In addition, the 
van der Waals
interaction gives rise to blockade phenomena; i.e., 
once a Rydberg state is
excited the excitation of an additional Rydberg atom is strongly suppressed in
the surrounding area characterized by a blockade radius $a_{R}$.  The correct
description of this property is obtained by the pair-correlation function 
%
$   g_{2}({\bf r}_{i}- {\bf
r}_{j})= 
\langle \p_{ee}^{(i)} \p_{ee}^{(j)} \rangle
/f_{R}^{2}
$, 
%
where $g_2({\bf r})$ vanishes in the blockaded region for $|{\bf r}| \ll a_{R}$,
while at large distances $|{\bf r}| \gg a_{R}$ the correlation disappears and
$g_2({\bf r}) = 1$. The transition from a strong suppression to the uncorrelated
regime is very sharp due to the van der Waals repulsion \cite{Robicheaux2005},
and the pair-correlation function is well described by a step function $
g_2({\bf r}) = \Theta\left(|{\bf r}|- a_R\right)$. Then, the mean-field theory
is obtained by replacing the microscopic interaction 
by the mean interaction of the surrounding atoms
\begin{equation}
  \label{eq:approx}
  \p_{ee}^{(i)}\p_{ee}^{(j)} \approx \left[\p_{ee}^{(i)}f_{R}+\p_{ee}^{(j)}f_R-f_R^2\right] g_2({\bf
    r}_{i}-{\bf r}_{j}),
\end{equation}
which neglects the quadratic fluctuations around the mean field and reduces the
Hamiltonian to a sum of single site Hamiltonians, $\h = \sum_{i} \h_{\rm
\scriptscriptstyle MF}^{(i)}$. In the scaling regime with $\alpha \ll 1$, the
number of atoms in the blockaded regime is large, i.e., $a_{R}^{3} n \gg 1$,
which allows us to replace the summation over the surrounding atoms $j$ by an
integral over space with a homogeneous atomic density $n$. Then, we obtain the
Hamiltonian for the $i^{\mathrm{th}}$ atom,
\begin{equation}
  \label{eq:heff} \frac{\h_{\rm \scriptscriptstyle MF}^{(i)}}{E_0} =
\frac{\alpha}{2} \s_x^{(i)} + \frac{4 \pi}{3} \frac{f_{R}}{n a_R^3}
\p_{ee}^{(i)} -\frac{2\pi}{3}\frac{f_R^2}{na_R^3}  = {\bf h}\cdot \hat{{\boldsymbol
\sigma}}^{(i)} + h_{0}.
\end{equation}
with the characteristic energy scale $E_{0} = C_{6} n^{2}$. Note that
in equilibrium the correlation function $g_2({\bf r})$ satisfies the
normalization condition $n f_R \int d{\bf r} \left[1-g_2({\bf
    r})\right] = 1$, which provides the relation $a_{R} = ( 3/4 \pi
f_{R} n)^{1/3}$. The effective Hamiltonian is equivalent to a spin in
a magnetic field ${\bf h}$ with $h_{x} = \alpha/2 $ and $h_{z}=8\pi^2
f_{R}^2/9$, and a constant energy offset $h_0=h_{z}(1-f_R)$. Using a
spin rotation, we can diagonalize the Hamiltonian $\h_{\rm
  \scriptscriptstyle MF}^{(i)} = h \s_{z'}^{(i)}+h_0 $ with $h =
\sqrt{h_x^2+h_z^2}$. Here, $\s_{z'}^{(i)} = \cos \theta \s_{z}^{(i)} +
\sin \theta \s_{x}^{(i)}$ denotes the Pauli matrix in the new basis
with the rotation angle $\tan \theta = h_{x}/h_{z}$. Within the new
basis the stationary equilibrium density matrix $\dop^{(i)}$ is
diagonal.  Its entries dependent on the experimental realization of
the system: (i) for an adiabatic switching on of the Rabi frequency,
the system remains in the ground state and the density matrix reduces
to the lowest energy state of $\h_{\rs{MF}}^{(i)}$. (ii) on the other
hand, for a sudden switching on of the Rabi frequency, the density
matrix is determined by energy conservation: the energy of the
equilibrated state is equal to the energy of the initial state, i.e.,
$ \tr{\h_{\rm \scriptscriptstyle MF}^{(i)} \dop^{(i)}} = \langle\psi |
\h | \psi \rangle / N= 0$ (here, $N$ denotes the total number of
particles). It can be checked that the scaling exponent remains the
same in both cases, indicating that the equilibrium state in the
second case is close to the ground state.  In the following, we focus
on a sudden switching on of the Rabi frequency. The mean-field
solution for the equilibrium state reduces to $\dop^{(i)}= \left[1
  -(h_0/h)\s_{z'}^{(i)}\right]/2$, and a transformation into the
original coordinates yields the self-consistency relation
\begin{equation}
  f_{R} = \langle \hat{P}_{ee}^{(i)} \rangle =
  \frac{1}{2}\frac{(4\pi)^4f_R^5+(9\alpha)^2}{(4\pi
    f_R)^4+(9\alpha)^2}.
\end{equation}
The solution in the limit $\alpha \ll 1$ provides the critical exponent $\nu =
2/5$ with the  prefactor $c = (9/16 \pi^2)^{2/5}$.

The same mean-field analysis can also be performed for arbitrary dimensions
$d\leq 4$.  The main modification is that the dimensionless parameter
$\alpha_{d}=\hbar \Omega/C_{6} n^{6/d}$ exhibits a dependence on the dimension
$d$ and also $h_{z}$ obeys the modified scaling $h_{z} \sim f_{R}^{6/d}$.  The
general result provides the scaling exponent $\nu_{d}= 2 d/(12+d)$.

\begin{figure}
  \includegraphics[width= 0.7\columnwidth]{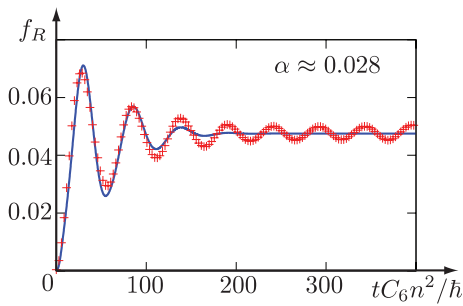}
  \caption{(color online). Numerical integration of the full
    Hamiltonian (crosses) for $\alpha = 1/36$, $N = 60$ and averaged
    over 50 different initial conditions.  The remaining oscillations
    at large times are finite size effects.  The solution of the
    master equation is shown as solid line.}  \label{fig:time}
\end{figure}

We now compare the above mean-field solution to numerical studies of
the time evolution with the full Hamiltonian Eq.~(\ref{eq:H}). We
place $N$ atoms randomly into a box of volume $V$ having periodic
boundary conditions. The dimension of the full Hilbert space grows
exponentially like $2^N$; therefore, the exact dynamics can only be
calculated for a relatively small number of atoms
\cite{Robicheaux2005,Hernandez2008}.  However, the strong van der
Waals repulsion suppresses the occupation probabilities of many basis
states, which allows us to significantly reduce the Hilbert space: for
each basis state we compute the van der Waals energy and remove the
state if its van der Waals energy is larger than a cutoff energy
$E_C$.  This reduction leads for $N = 100 $ to approximately $10^6$
relevant basis states compared to the $10^{30}$ basis states of the
full Hilbert space.  Convergence has been checked by increasing
$E_C$. We have used a fourth-order Runge-Kutta scheme for performing
the numerical integration of the Schr\"odinger equation. The
time-dependence of the Rydberg fraction $f_R(t)= \sum_{i} \langle
P_{ee}^{i} \rangle/N$ is shown in Fig.~\ref{fig:time} (crosses), for
an average of 50 different uniformly distributed random initial
conditions with $N=60$ and $\alpha = 1/36$. We find a clear saturation
of the excited Rydberg fraction $f_{R}$; the remaining oscillations at
large times are further suppressed for larger system sizes, and
therefore represent a finite size effect.

\begin{figure}[ht]
  \includegraphics[width=\columnwidth]{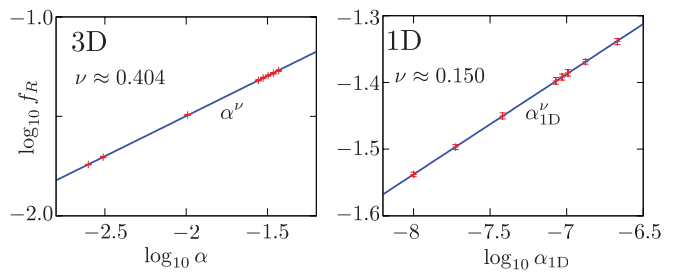}
 \caption{(color online). Numerical results for the saturated Rydberg
   fraction $f_{R}$ in a 3D and a 1D setup: the system exhibits an
   algebraic behavior $f_{R}\sim \alpha^\nu$ with $\nu\approx 0.404$
   and $\nu_{1D}\approx0.150$.}

\label{fig:nsat}

\end{figure}

In order to investigate the scaling behavior of the saturated Rydberg fraction
$f_{R}$ we have taken an average over the Rydberg fraction $f_R(t)$ for times
$250 \leq t C_6 n^2/\hbar \leq 400$ for 50 different random initial
conditions.  The dimensionless parameter $\alpha$ has been varied by
changing the number of atoms from $N = 52$ to $N = 100$, and by
changing $C_6$ from $0.01$ to $0.04$.  The scaling behavior is shown
in Fig.~\ref{fig:nsat}. The data show a power law dependence according
to $f_{R} = c\alpha^{\nu}$.  The fit to the numerical data provides an
exponent $\nu = 0.404$, which is in very good agreement with the
result derived in the mean-field theory.  For the onedimensional case
the numerically obtained critical exponent is $\nu_{\rm
  \scriptscriptstyle 1D} = 0.150$. Surprisingly, this value is again
very close to the mean-field prediction, indicating that the van der
Waals interaction strongly suppresses quantum fluctuations.

Finally, we are interested in a description of the time evolution and derive a
master equation with the mean-field solution as its stationary state. 
The natural mechanisms for 
the equilibration into a stationary state are the residual
interactions between the atoms, which go beyond the mean-field
description. We write the exact Hamiltonian (\ref{eq:H}) as a sum of
the mean-field terms and the remaining fluctuations $ \h = \sum_{i}
\h_{\rm \scriptscriptstyle MF}^{(i)} + \Delta \h$.
%
%
The derivation of the master equation from the microscopic Hamiltonian
uses the time-convolutionless projection operator method with an
extended projection operator \cite{Breuer2007}: we select a single
site $i$, which will play the role of the system with the Hamiltonian
$\h_{\rm \scriptscriptstyle MF}^{(i)}$, while the surrounding atoms
act as the bath coupled to the system state by the Hamiltonian $\Delta
\h$.  The role of the pair-correlation function is to enforce the
blockade regime.  Here, we
assume the pair-correlation function to be fixed during the time
evolution. Then, its influence is well accounted for by expressing the
remaining interactions as $\Delta \h = \sum_{i<j} g_2({\bf r}_{i}~-~{\bf
  r}_{j})\,C_6/|{\bf r}_{i}-{\bf
  r}_{j}|^{6}\left(\p_{ee}^{(i)}-f_{R}\right)\left( \p_{ee}^{(j)} -
f_{R}\right)$.
The projection operator consistent with our mean-field theory, i.e.,
$\mathcal{P}\Delta \h \dop = 0$ reduces to $ \mathcal{P}\dop =
\bigotimes_{i} \dop^{(i)}$. Here, $\dop^{(i)}$ denotes the reduced
density matrix defined by the partial trace $ \dop^{(i)} = {\rm
  Tr}_{i}\left\{ \dop\right\}$, which performs the trace over all
atomic states except of the $i^{\mathrm{th}}$ atom.  Next, it is
useful to express the operator $\hat{P}_{ee}^{(i)} =
\aop{-\omega_{0}}^{(i)} +\aop{0}^{(i)}+\aop{\omega_{0}}^{(i)} $ in
terms of the projections onto the eigenstates of
$\h_{\rm \scriptscriptstyle MF}^{(i)}$, where $\hbar\omega_{0}= 2h$
denotes the energy splitting of the mean-field Hamiltonian. 
Introducing the interaction picture with respect to $\sum_{i} \h_{\rm
  \scriptscriptstyle MF}^{(i)}$ the interaction Hamiltonian
$\h^{(i,j)}_{\rm \scriptscriptstyle int}(t)$ within the rotating wave
approximation reduces to
\begin{equation}
   \h^{(i,j)}_{\rm \scriptscriptstyle int}(t)  = \sum_{\omega= 0, \pm\omega_{0}}
\frac{C_{6}}{|{\bf r}_{i}-{\bf r}_{j}|^6}
  \aop{\omega}^{(i)}\otimes\bop{\omega}^{(j)\dagger}.
\end{equation}
The terms $\aop{\pm \omega_{0}}$ describe the exchanges of an excitation between
the system $\dop^{(i)}$ and the surrounding bath and is
relevant for the equilibration of the time evolution, while the terms with
$\aop{0}$ account for a dephasing.  The second order time-convolutionless master
equation \cite{Breuer2002,Breuer2007} for the reduces density matrix
$\dop^{(i)}(t)$ takes the form
\begin{displaymath}
\dod{}{t} \dop^{(i)} = \sum_{\omega,\omega'}
\gamma_{\omega\omega'}\left(\aop{\omega}^{(i)} \dop^{(i)}
\aop{\omega'}^{(i) \dagger} -
\frac{1}{2}\anticom{\aop{\omega'}^{(i)\dagger}\aop{\omega}^{(i)}}{\dop^{(i)}}\right),
\end{displaymath}
with the rates
\begin{eqnarray}
  \gamma_{\omega\omega'}& = &\frac{512\pi^4 n t}{27\hbar^2}\frac{1}{a_{R}^{9}}
\langle\bop{\omega'}^{(i)}\bop{\omega}^{{(i)}\dagger}\rangle.
\end{eqnarray}
Here, we have again used the translation invariance of the system in
the critical region $\alpha \ll 1$ allowing us to replace the average
over the $j^{\mathrm{th}}$ atom, with the local density matrix
$\dop^{(i)}(t)$. The master equation is a highly nonlinear equation
for the local density $\dop^{(i)}(t)$, where at each time step the
mean-field $f_{R}={\rm Tr} \{\dop^{(i)}\hat{P}_{ee}^{(i)}\}$ and the
rates $\gamma_{\omega,\omega'}$ have to be determined. We would like
to stress that the master equation conserves energy with the above
mean-field solution as a stationary state.  The master equation can be
efficiently solved numerically after a transformation back into the
Schr\"odinger picture. In Fig.~\ref{fig:time} the solution of the
master equation (solid line) is compared to the full dynamics
(crosses). The pair-correlation function strongly influences the
decoherence rates $\gamma_{\omega,\omega'}$.  As the pair correlation
is not determined self-consistently within our approach, we apply once
a single fit for the effective Rabi frequency and the saturated
Rydberg fraction to account for these corrections. Then, we find
perfect agreement between the full dynamics and the solution of the
master equation, see Fig.~\ref{fig:time}; the deviations at large
times are accounted to the finite size effects of the full dynamics.
The characteristic time scale of the oscillations is determined by the
collective Rabi frequency $ \sqrt{N_{\rm \scriptscriptstyle
    block}}\Omega$, with $N_{\rm \scriptscriptstyle block}\sim n
a_{R}^{3}$ the number of atoms within the blockaded volume of the van
der Waals interaction.

In conclusion, we have established a universal scaling behavior in
strongly interacting Rydberg gases due to the existence of a second
order quantum phase transitions. It remains an open question, whether
it is experimentally possible to reach the crystalline phase.


We acknowledge discussions with F. Robicheaux. The work was supported
by the Deutsche Forschungsgemeinschaft (DFG) within SFB/TRR 21 and by
Pf381/4-1.


\begin{thebibliography}{10}

\bibitem{Fisher1998}
M.~E. Fisher,
\newblock Rev. Mod. Phys. {\bf 70}, 653 (1998).

\bibitem{Anderson1998}
W.~R. Anderson, J.~R. Veale, and T.~F. Gallagher,
\newblock \prl {\bf 80}, 249 (1998).

\bibitem{Singer2004}
K.~Singer et~al.,
\newblock \prl {\bf 93}, 163001 (2004).

\bibitem{Tong2004}
D.~Tong et~al.,
\newblock \prl {\bf 93}, 063001 (2004).

\bibitem{Vogt2006}
T.~{Vogt} et~al.,
\newblock \prl {\bf 97}, 083003 (2006).

\bibitem{Heidemann2007}
R.~Heidemann et~al.,
\newblock Phys. Rev. Lett. {\bf 99}, 163601 (2007).

\bibitem{Raitzsch2008}
U.~Raitzsch et~al.,
\newblock Phys. Rev. Lett. {\bf 100}, 013002 (2008).

\bibitem{Heidemann2008}
R.~Heidemann et~al.,
\newblock Phys. Rev. Lett. {\bf 100}, 033601 (2008).

\bibitem{Johnson2008}
T.~A. Johnson et~al.,
\newblock Phys. Rev. Lett. {\bf 100}, 113003 (2008).

\bibitem{Cubel2005}
T.~Cubel et~al.,
\newblock Phys. Rev. A {\bf 72}, 023405 (2005).

\bibitem{Mohapatra2007}
A.~K. Mohapatra, T.~R. Jackson, and C.~S. Adams,
\newblock Phys. Rev. Lett. {\bf 98}, 113003 (2007).

\bibitem{Reetz-Lamour2008}
M.~Reetz-Lamour et~al.,
\newblock New J. Phys. {\bf 10}, 045026 (2008).

\bibitem{Robicheaux2005}
F.~{Robicheaux} and J.~V. {Hern{\'a}ndez},
\newblock \pra {\bf 72}, 063403 (2005).

\bibitem{Stanojevic2008}
J.~{Stanojevic} and R.~{C{\^o}t{\'e}},
\newblock arXiv:0801.2406  (2008).

\bibitem{Ates2007}
C.~Ates et~al.,
\newblock Phys. Rev. A {\bf 76}, 013413 (2007).

\bibitem{Stanojevic2008a}
J.~Stanojevic and R.~C\^ot\'e,
\newblock arXiv:0801.2396  (2008).

\bibitem{Gallagher1994}
T.~F. Gallagher,
\newblock {\em Rydberg Atoms} (Cambridge University Press, Cambridge, 1994).

\bibitem{Singer2005}
K.~{Singer} et~al.,
\newblock J. Phys. B {\bf 38}, S295 (2005).

\bibitem{Robicheaux2005a}
F.~Robicheaux,
\newblock J. Phys. B {\bf 38}, S333 (2005).

\bibitem{Amthor2007}
T.~Amthor et~al.,
\newblock Phys. Rev. Lett. {\bf 98}, 023004 (2007).

\bibitem{Hernandez2008}
J.~V. Hern\'{a}ndez and F.~Robicheaux,
\newblock J. Phys. B {\bf 41}, 045301 (2008).

\bibitem{Breuer2007}
H.-P. Breuer,
\newblock Phys. Rev. A {\bf 75}, 022103 (2007).

\bibitem{Breuer2002}
H.-P. Breuer and F.~Petruccione,
\newblock {\em The {T}heory of {O}pen {Q}uantum {S}ystems} (Oxford University
  Press, Oxford, 2002).

\end{thebibliography}

\end{document}